\documentstyle[preprint,aps,prl,psfig]{revtex}
\tighten
\draft
\begin{document}

\draft

\title{Amorphous silica between confining walls and under shear:
       a computer simulation study}
\author{J\"urgen Horbach and Kurt Binder}
\address{Institut f\"ur Physik, Johannes Gutenberg--Universit\"at,
Staudinger Weg 7, D--55099 Mainz, Germany}
\maketitle

\begin{abstract}
Molecular dynamics computer simulations are used to investigate a
silica melt confined between walls at equilibrium and in a steady--state
Poisseuille flow. 
The walls consist of point particles forming a rigid
face--centered cubic lattice and the interaction of the walls with
the melt atoms is modelled such that the wall particles have only a weak bonding to those
in the melt, i.e.~much weaker than the covalent bonding of a Si--O unit. 
We observe a pronounced layering of the melt near the
walls. This layering, as seen in the total density profile, has a very
irregular character which can be attributed to a preferred orientational
ordering of SiO$_4$ tetrahedra near the wall. On intermediate length
scales, the structure of the melt at the walls can be well distinguished
from that of the bulk by means of the ring size distribution. Whereas
essentially no structural changes occur in the bulk under the influence
of the shear fields considered, strong structural rearrangements in the
ring size distribution are present at the walls as far as there is a
slip motion. For the sheared system, parabolic velocity profiles are 
found in the bulk region as expected from hydrodynamics and the values 
for the shear viscosity as extracted from those profiles are in good 
agreement with those obtained in pure bulk simulations from the appropriate 
Green--Kubo formula.
\end{abstract}


\section{Introduction}
\label{sec1}

The transport properties of atomic fluids that move through nanoscopic
slits between solid walls has been an issue of constant interest
over the past decades~\cite{evans90,hess97,churaev00,mueser02a,mueser02b}. Many of
the simulation studies investigating such fluids in Couette or
Poisseuille flow have considered simple monoatomic systems in which
the particles interact via Lennard--Jones (LJ) type forces (see, e.g.,
\cite{thompson97,todd95,travis97,travis00}). One of the remarkable
results of these studies was that velocity profiles as predicted
by hydrodynamics can be observed even if the distance between the
confining walls is of the order of only ten molecular diameters. But
one has to keep in mind that the monoatomic LJ systems for which
this was found have typically a relatively small viscosity (of the
order of $10^{-4}$ to $10^{-3}$~Poise in Argon units) and furthermore, these systems
crystallize very easily if one decreases the temperature below the
melting point. An open question is what happens with more complicated
sheared fluids that are confined between walls, i.e.~especially liquids
that have a structure that is not similar to the local structure of
a closed--packed system of hard spheres like the aforementioned LJ
systems. One class of more complex materials for which flow properties
in confinement have already been studied to some extent are polymer
melts~\cite{onuki97,yamamoto00,kroger93,kroger00,aust99,kreer01,varnik02}.
Very recently it was shown~\cite{varnik02} for such a system that 
its properties under shear can be well understood within a hydrodynamic 
description.

In this paper we are interested in amorphous silica (SiO$_2$) which is
the paradigm of a network forming liquid. We consider a SiO$_2$ melt that
is confined between two crystalline walls and on which a gravitational--like
field is exerted in order to yield a stationary Poisseuille flow.
It is difficult to address this problem in a simulation because model
potentials of silica contain long--ranged Coulomb terms~\cite{poole} for which the
corresponding force calculations become very time--consuming for a three
dimensional system in which periodic boundary conditions are considered
only in two directions. This stems from the fact that the Fourier
part of the Ewald sums can no longer be calculated by a single loop
over the number of particles $N$ as in the case of periodic boundary
conditions in three dimensions, but one has to compute a double loop that
scales with $N^2$~\cite{parry75,parry76,heyes94,leeuw82}. Therefore,
very recently, many efforts were undertaken to develop alternative
methods~\cite{widmann97,strebel01,arnold02,joannis02} for the summation
of ``two--dimensional'' Coulomb interactions. But most of these new approaches
are not established yet and extensive tests that demonstrate that their
accuracy is comparable to the ``brute force'' $N^2$--methods are still lacking.
Moreover, differently from monoatomic LJ systems, silica is a good 
glassformer that exhibits a slow dynamics even above the melting 
temperature. Note that at the experimental melting temperature, 
$T=2000$~K, amorphous silica has a viscosity of about $10^7$~Poise~\cite{mazurin}.  
Systems with such a high viscosity cannot be equilibrated nowadays 
by means of a molecular dynamics computer simulation.

The structure of silica is that of a disordered tetrahedral network in
which SiO$_4$ tetrahedra are connected to each other in that they share
an oxygen atom at their corner. Extensive simulation studies of free
silica surfaces, that used both classical molecular dynamics as well as
Car--Parrinello simulations~\cite{roder01,mischler02} gave evidence for
a different packing of the SiO$_4$ tetrahedra at the surface from that
in the bulk. This different packing is manifested very clearly in the
ring structure (a ring is a closed loop formed by consecutive Si--O elements).
At the surface there is a relatively high probability to
find short rings such as two-- and three--membered rings whereas
in the bulk the occurence of such rings is very rare even at relatively
high temperatures. Note that the occurrence of short rings on
a free silica surface has been confirmed experimentally by means of
atomic force microscopy~\cite{poggemann01}. A detailed understanding
of the structure near the wall is important to give insight into the
microscopic origin of the wall--fluid boundary conditions if a flow field
is introduced. This might be especially important in the case of silica
since it does not form a closed--packed structure and thus the behavior
could be very different from that of a hard--sphere like system. Indeed we
show below that there is a rearrangement in the ring distribution near
the walls if the shear is strong enough.

An important issue for the understanding of a confined fluid under
shear are the boundary conditions that enter in the hydrodynamic
description in order to take into account the interactions of the fluid
with the wall. In principle, these boundary conditions can be derived
from the microscopic equations of motion for the fluid and the wall
particles. But in practice one often introduces as an {\it ad hoc}
assumption stick boundary conditions. However, it has been found in
simulations of simple fluids that there is often a slip motion of the
fluid at the wall~\cite{barrat99a,barrat99b}. So in general one has to
consider slip--stick boundary conditions for a hydrodynamic analysis
of fluid flow between walls. Bocquet and Barrat~\cite{bocquet94} have
shown in a microscopic theory, in which they have related quantities
like the slip length with equilibrium time correlation functions via
Green--Kubo formulas, that slip occurs if the walls do not provide a
strong corrugation. E.g., a large slip effect is observed if the fluid
partially wets the solid walls.  In a recent simulation of Sokhan {\it
et al.}~\cite{sokhan01} it has been even demonstrated for the fluid
flow through a carbon tube that slip--stick boundary conditions are the
generic case if one uses realistic parameters in the model potential for
the wall--fluid interactions.  It is one of the issues of this paper to
characterize the structure of the silica melt near the wall and to see
how the structure is related to the observed boundary conditions.

Another issue that we address in this paper is to discuss to what
extent the velocity profiles from our simulations can be used to
estimate the shear viscosity $\eta$ of a highly viscous fluid such as
silica. An alternative method is to calculate $\eta$ from the appropriate
Green--Kubo formula~\cite{hansen} which expresses $\eta$ as the time
integral over a stress--stress autocorrelation function. We applied
the latter approach recently in extensive bulk simulations of silica
in which we were able to determine $\eta$ in the temperature range
6100~K~$\ge T \ge$~3000~K whereby at $T=3000$~K the visosity is of the
order of $60$~Poise~\cite{horbach99}. In the present paper we use the
latter results as a reference to see how reliable the estimates of $\eta$
from our non--equilibrium simulations are. Such a reference is important
since it is not obvious that the observed velocity profiles justify a
``hydrodynamic analysis'' keeping in mind that the distance between the
confining walls is on a nanoscopic scale.

The rest of the paper is organized as follows: In Sec.~\ref{sec2}
we describe how we model the walls and the silica melt, and we give
the details of the simulation.  Sec.~\ref{sec3} is then devoted to
the presentation of the results and eventually, in Sec.~\ref{sec4}
we summarize and discuss them.

\section{Model and Details of the Simulation}
\label{sec2}

A realistic simulation of silica confined between walls requires an
interaction potential that yields a reliable description of both
the bulk properties and the properties at the interface between
the walls and the silica fluid.  As far as the bulk properties are
concerned recent simulations have shown that the potential that was
proposed some years ago by van Beest {\it et al.}~\cite{beest90}
(the so--called BKS potential) is able to reproduce various
properties of amorphous silica very well, such as its structure,
its vibrational and relaxational dynamics, the static specific
heat below the glass transition temperature, and the conduction of
heat~\cite{horbach99,vollmayr96,koslowski97,taraskin97,horbach99b,horbach01,scheidler01,jund99,benoit00}.
Recent extensive simulations~\cite{mischler02} that used a combination
of a classical molecular dynamics simulation with the BKS potential
and {\it ab initio} simulations (Car--Parrinello molecular dynamics)
gave evidence that the BKS potential yields also a fair description of
the structure of free amorphous silica surfaces, and thus, it can be
expected that it also reproduces the main features of the structure of
amorphous silica near a wall.

The functional form of the BKS potential is given by
\begin{equation}
\phi_{\alpha \beta}(r)=
\frac{q_{\alpha} q_{\beta} e^2}{r} +
A_{\alpha \beta} \exp\left(-B_{\alpha \beta}r\right) -
\frac{C_{\alpha \beta}}{r^6}\quad \alpha, \beta \in
[{\rm Si}, {\rm O}],
\label{eq1}
\end{equation}
where $r$ is the distance between the ions of type $\alpha$ and $\beta$.
The values of the constants  $q_{\alpha}, q_{\beta}, A_{\alpha
\beta}, B_{\alpha \beta}$, and $C_{\alpha \beta}$ can be found in
Ref.~\cite{beest90}. For the sake of computational efficiency the short
range part of the potential was truncated and shifted at 5.5~\AA. This
truncation has also the benefit to improve the agreement between the
density of the amorphous glass at low temperatures as determined from
the simulation with the experimental value. For the calculation of the
long--ranged Coulomb interactions we used the method first introduced by
Parry~\cite{parry75,parry76} which is a straightforward generalization
of the three--dimensional Ewald sums with periodic boundary conditions
in all three directions.  In the latter case the Fourier sum requires
only a loop over the $N$ particles of the system which leads, together
with the real space sum, to an effective computational load that scales
as $N^{1.5}$~\cite{frenkel}.  In contrast to that in the Parry--Ewald
sums for systems with a quasi--two--dimensional geometry the Fourier and
the real space part have to be computed by a double loop over all pairs
of particles.  However, we preferred to use Parry--Ewald sums rather than
alternative faster methods~\cite{widmann97,strebel01,arnold02,joannis02}
since it provides accurate results and it is easy to implement.

The walls were not constructed as to model a particular material but
rather to be a generic surface that can be simulated conveniently. Each
wall hence consisted of 563~point particles forming a rigid face--centered
cubic lattice with a nearest--neighbor distance of $2.33$~\AA. These
point particles interact with the atoms in the fluid according to a 12--10
potential,
\begin{equation}
  v(r) = 4 \epsilon \left[
	 (\sigma / r)^{12} - (\sigma / r )^{10} \right] \  ,
  \label{wall}
\end{equation}
with $\sigma=2.1$~\AA, $\epsilon=1.25$~eV, $r$ being the distance
between a wall particle and a Si or O atom. Note that we choose the same
potential between the wall atoms and both Si and O because we wish to
simulate the generic effects of confinement, separated from possible
additional effects due to surface enrichment of one component caused
by the difference in surface forces, which may be a problem in many
real systems. The great advantage of simulation is that by idealizing
certain aspects of a problem a much better physical understanding of
consequences of particular interactions may be achieved. In this spirit,
a wall potential has been ``invented'' that keeps the slip small (when we apply
a constant force in $x$--direction to the particles causing a corresponding
flow) and provides at the same time a weak bonding of the wall
particles to those of the melt such that no covalent bonds
of wall particles with Si or O atoms are formed. In order fulfil these two 
``requirements'' the $r^{-10}$--term in Eq.~(\ref{wall}) has turned
out to be more suitable than a $r^{-6}$--term. 

The simulation box had linear dimensions $L_x=L_y=23.066$~\AA~in the
directions parallel to the walls (in which also periodic boundary
conditions were applied), and $L_z=31.5$~\AA~in the direction
perpendicular to the walls. Thus, $N=1152$~atoms (384 Si atoms and 768
O atoms) were contained in the system to maintain a density around
$2.3$~g$/$cm$^{3}$ which is close to the experimental one at zero
pressure~\cite{mazurin}.  In the Parry--Ewald sums the parameter
$\alpha$ and the cutoff wave--vector for the Fourier part were
chosen to $0.265$ and $6$, respectively~\cite{frenkel}. The equations of motion
were integrated with the velocity form of the Verlet algorithm with
a time step of 1.6~fs. All runs were done in the $NVT$ ensemble
whereby the temperature was kept constant by coupling the fluid to a
Nos\'e--Hoover thermostat~\cite{nose83,nose84,hoover85}.  Whenever an
external gravitational--like field was switched on, the thermostat was
only applied in the two directions perpendicular to the flow, i.e.~in
$y$-- and $z$--direction, whereas otherwise the thermostat was used in
all three directions.  Bocquet and Barrat~\cite{barrat99b} also embarked
on this strategy to provide a correct adjustment of the temperature when
the system is sheared.

We investigated the three temperatures $T= 5200$~K, $T=4300$~K,
and $T=3760$~K at which we first fully equilibrated the system
for $29$~ps, $65$~ps, and 122~ps, respectively. At $T= 5200$~K and
$T=3760$~K we continued with additional runs over 164~ps and 490~ps,
respectively, from which we analyzed the equilibrium structure.  Then we
switched on a gravitational acceleration field of strength $a_{{\rm
e}}=9.6$~\AA$/$ps$^2$ that was coupled to the mass of the particles.
We mention that although the masses of oxygen and silicon differ by about
a factor of two ($m_{{\rm O}}=15.9994$~u, $m_{{\rm Si}}=28.086$~u) we
obtained identical velocity profiles for both species (see below), and
this is due to the fact that the chemical bonding force between oxygen
and silicon atoms is much stronger than the imposed acceleration field.
With the acceleration field runs were made over $736$~ps, $1.23$~ns, and
$3.27$~ns at $T=5200$~K, 4300~K, and 3760~K, respectively.  In addition
at $T=5200$~K we did a run over $1.72$~ns with field strength $a_{{\rm
e}}=3.8$~\AA$/$ps$^2$.  During these runs we stored the positions
and velocities of the particles every $0.16$~ps (i.e.~every 100 time
steps). For the calculation of the steady state properties we used only
those data that were obtained after two times the time that was used
before for the equilibration.  The total amount of computer time spent
for these simulations was 16~years of single processor time on a Cray T3E.

\section{Results}
\label{sec3}

In this section we present the simulation results for the silica
melt between confining walls. We characterize the structure of the melt
and show how it changes if one switches on a gravitational field, i.e.,
if a Poisseuille flow is present in the melt. By analyzing the velocity
profiles we discuss to what extent their behavior can be understood within a
hydrodynamic description.

The snapshot, Fig.~\ref{fig1}, illustrates on what length scales the walls
affect the structure of the melt.  It shows a part of a configuration
at $T=3760$~K, namely the part that is within a distance less than
$8$~\AA~away from one of the walls. One can clearly identify a layering
of SiO$_4$ tetrahedra which becomes less pronounced if one moves away
from the wall.  Note that at $T=3760$~K $95$\% of the silicon atoms
are four--fold coordinated by oxygen atoms and 96.5\% of the oxygen
atoms are two--fold coordinated by silicon atoms (a silicon and an
oxygen atom are defined as neighbors if their distance is less than
$2.35$~\AA~corresponding to the first minimum of the pair correlation
function for Si--O). The rest of the atoms form defects, i.e.~silicon atoms
that are three-- or five--fold coordinated by oxygen atoms and oxygen
atoms that are one-- or three--fold coordinated by silicon atoms. Thus, since
less than 5\% of the atoms are involved in such defects it makes sense
to characterize the structure also at the relatively high temperature
$T=3760$~K as a disordered tetrahedral network where SiO$_4$ tetrahedra
are connected via bridging oxygens (i.e.~every oxygen atom belongs to
two tetrahedra).

The layering of the SiO$_4$ network near the wall can be quantified
by means of the density profile which is plotted in Fig.~\ref{fig2}a
across half of the film~\cite{foot1} for all atoms and for the oxygen
and silicon atoms only.  In contrast to typical density profiles in
simple monoatomic liquids the oscillations of the total profile, which
indicate the layering near the walls, do not have a regular character. But
this irregular behavior can be understood if one looks at the partial
profiles for oxygen and silicon. The explanation that emerges is as
follows: As we mentioned before confined fluid SiO$_2$ maintains the
irregular network SiO$_4$ tetrahedra as in the bulk.  At the same time
the tetrahedra have to fill the available space and have to satisfy the
bonding energy to the wall.  To this end, it is necessary to align the
tetrahedra adjacent to the wall such that a two--dimensional plane forms
which contains three out of the four oxygen atoms of a tetrahedron,
as well as the silicon atom at their center (slightly further away
from the wall), while the fourth oxygen atom of the tetrahedron has to
be further away from the wall for geometrical reasons: this causes the
second peak of the oxygen distribution. Thus, the walls have a tendency
to ``orient'' the network of coupled SiO$_4$ tetrahedra in the fluid
locally. It is evident that the oscillations in the local density of both
silicon and oxygen are rather regular, like a damped cosine function,
but the wavelength and phase of both cosine functions are different:
their superposition causes then the rather irregular layering structure
of the SiO$_2$ total density. We expect that similar effects also occur
in many other associating molecular fluids confined between walls if
the wall--fluid interaction is weak enough that it does not affect the
chemical ordering in the fluid as it is the case in our system.

For $z > 8$~\AA~the total density profile shows only small oscillations
around a constant value of 2.3~g$/$cm$^3$ which is an indication for bulk
behavior. Thus, keeping in mind that the density profile is symmetric,
the bulk in our system seems to extend in $z$ direction from about
$8$~\AA~to $23.5$~\AA, i.e.~it has a width of about 15.5~\AA. However,
Fig.~\ref{fig2}a shows that the deviations from a constant density in the
latter region are not due to the statistics: The partial density profiles
exhibit still oscillations also in the middle of the film that one would
not expect in a pure isotropic bulk system.  But these oscillations do
not affect the structural quantities that we show below, i.e.~the pair
correlation functions and the distributions of angles and rings, since
these quantities essentially do not differ from the corresponding ones
of a pure bulk simulation at the same temperature. Therefore, we consider
in the following the region defined by $8$~\AA~$\le z \le 23.5$~\AA~as
the bulk.

The behavior of the total density profiles at the temperatures $T=5200$~K
and $T=3760$~K in equilibrium and with an external force with an
acceleration of $9.6$~\AA$/$ps$^2$ can be seen in Fig.~\ref{fig2}b. In
equilibrium the effect of decreasing temperature on the oscillations near
the wall is an increase of the peak heights that is accompanied with a
smaller value of $\rho(z)$ at the minima between the peaks. Thus, the
layering becomes more pronounced if one decreases the temperature. If
one switches on the gravitational--like field the effect is similar to an
increase of the temperature. In the bulk region the density profiles are
not very sensitive to a variation of temperature and/or the presence
of the external force. Within the accuracy of our data the same
value of about $2.3$~g$/$cm$^3$ is reached for all four cases under
consideration. However, this does not exclude of course the possibility
of a dramatic change in the local structure, but as we will see in the
following, the local structure is not strongly affected by the considered
external forces.

Quantitities that provide information on the microscopic interparticle
distances are the partial pair correlation functions $g_{\alpha \beta}(r)$
which are proportional to the probability of finding an atom of type
$\alpha$ at a distance $r$ from an atom of type $\beta$. In an isotropic
system the functions $g_{\alpha \beta}(r)$ are normalized by a phase
factor $4 \pi r^2$~\cite{hansen}. However, for the case that boundaries
are present this factor has to be modified in that it is then determined
only by that part of the surface $4 \pi r^2$ of a sphere around a particle
that fits into the system. Thus, the normalization factor depends on the
distance of the particles from the walls in $z$--direction.  In order
to identify the differences of $g_{\alpha \beta}(r)$ close to the wall
from those of the bulk we define a bulk region for $8$~\AA$\le z \le
23.5$~\AA~and wall regions for distances less or equal $3$~\AA~from the
wall (containing respectively the first oxygen and silicon layer, see
Fig.~\ref{fig2}a). The latter wall layer is denoted in the following
by WL.  The $g_{\alpha \beta}(r)$ are calculated only for particles
within the bulk or the WL whereby the aforementioned normalization
factors are determined such that also the boundaries of the bulk
region and the WL that are located within the fluid film are treated
as real boundaries. Fig.~\ref{fig3}~shows $g_{\alpha \beta}(r)$ for
the different correlations at $T=3760$~K, each at equilibrium in the
bulk and in the WL, as well as in the WL under the acceleration field
$a_{{\rm e}}=9.6$~\AA/ps$^2$. We have not included the curves for the
bulk with an external field because they essentially coincide with the
equilibrium curves. This means that for the field strength $a_{{\rm
e}}=9.6$~\AA/ps$^2$ the structure as reflected in $g_{\alpha \beta}(r)$
is not affected in the bulk by the gravitational field.  From this we
can consider the latter field strength as a small disturbance, and thus
we can expect that the system can be treated as a Newtonian fluid in the
bulk. In contrast to this at the walls there are significant changes in
$g_{\alpha \beta}(r)$ if one switches on the external field in that the
peaks become broader and their heights decrease slightly. These changes
are most pronounced in $g_{{\rm OO}}(r)$ which is reasonable since the
very first layer at the wall is formed by oxygen atoms.

The differences in $g_{\alpha \beta}(r)$ at the WL and in the bulk
reflect the different packing of the tetrahedral network in both regions.
In $g_{{\rm SiO}}(r)$ the second and third peaks are shifted to smaller
distances in the bulk as compared to the WL. Similar effects are even more
pronounced in $g_{{\rm OO}}(r)$ and $g_{{\rm SiSi}}(r)$.  In these
functions one can see that the fcc lattice of the wall affects the
structure on length scales of next--nearest Si--Si and O--O neighbors.
In $g_{{\rm OO}}(r)$ for the WL the periodicity of the peaks is
approximately given by the lattice constant $a=2.33$~\AA~of the fcc
lattice that is formed by the wall atoms. This periodicity appears
less pronounced in $g_{{\rm SiSi}}(r)$ since the Si atoms are not so
close to the wall as the O atoms and have thus more freedom to arrange
themselves. Moreover the first peak in $g_{{\rm SiSi}}(r)$ is at a
slightly smaller distance for the WL than for the bulk and this is due
to the denser packing of SiO$_4$ tetrahedra near the wall.

A further step towards the characterization of the local structure is to
consider correlations between triples of particles.  Simple quantities
to study such correlations are the distributions of angles between three
neighboring atoms.  Thereby, two atoms of type $\alpha$ and $\beta$ are
defined as neighbors if their distance is smaller than the position of the
first minimum in the corresponding $g_{\alpha \beta}(r)$.  The locations
of these minima are at 3.64~\AA, 2.35~\AA, and 3.21~\AA~for the Si--Si,
Si--O, and O--O correlations, respectively.  In Fig.~\ref{fig4}a we show
the angle distribution functions $P_{\alpha \beta \gamma}(\theta)$ for the
Si--Si--Si, Si--O--Si, and the O--Si--O angles in the bulk at equilibrium
and under shear and at the WL at equilibrium. The temperature in all cases
is 3760~K.  We see in the figure that in each of the $P_{\alpha \beta \gamma}(\theta)$ 
shown there are only small differences between the bulk curves for the 
equilibrium and the sheared case which again demonstrates
that the structure changes only slightly in the bulk region if a field
of strength $a_{{\rm e}}=9.6$~\AA$/$ps$^2$ is switched on.  There are
also only small differences in $P_{{\rm OSiO}}$ for the WL as compared
to the corresponding distributions in the bulk. Since the O--Si--O angle
is an intratetrahedral one this shows that the geometry of the SiO$_4$
tetrahedra is essentially the same at the wall and in the bulk. Remarkable
differences are in $P_{{\rm SiSiSi}}$ and $P_{{\rm SiOSi}}$ for the WL as
compared to corresponding functions for the bulk: First, the main peak in
$P_{{\rm SiOSi}}$ is shifted towards smaller angles which is due to the
denser packing of the tetrahedra in the WL. Secondly, there is a shoulder
in $P_{{\rm SiOSi}}$ around 100$^{\circ}$ which is less pronounced in the
WL and $P_{{\rm SiSiSi}}$ exhibits a secondary peak at about 60$^{\circ}$
which is more pronounced in the curve for the WL.  In Fig.~\ref{fig4}b
we compare the different $P_{\alpha \beta \gamma}(\theta)$ for the WL
at equilibrium and for the sheared system. As can be inferred from the
figure, in contrast to the bulk case there are changes in the angular
distributions in that all three distributions broaden significantly for
the system under shear. Moreover, the shoulder in $P_{{\rm SiOSi}}$ around
100$^{\circ}$ has about a six times larger amplitude in the latter case.

Recent studies of silica melts~\cite{roder01,mischler02} have shown
that the shoulder around 100$^{\circ}$ in $P_{{\rm SiOSi}}$ is due
to two--membered rings whereas the peak at about 60$^{\circ}$ in
$P_{{\rm SiSiSi}}$ corresponds to the presence of three--membered rings.
We look now for the latter ring sizes in our system and investigate the
distribution of rings in the network.  A ring is defined as follows:
One starts from any silicon atom and two of its nearest oxygen
neighbors. Then one counts the shortest consecutive sequence of Si--O
elements that connect the two latter oxygen atoms and this number is
then the ring length $n$. Thereby, the ring length $n=1$ corresponds
to a dangling bond, i.e., an O atom is only attached to one Si atom,
and in a two--membered ring ($n=2$) two silicon atoms share two oxygens.
Note that one can extract (indirectly) information on rings in SiO$_2$
by means of NMR experiments~\cite{stebbins}.

Fig.~\ref{fig5}a shows the ring distribution function $P(n)$ for
$T=3760$~K in the bulk and in two different wall layers denoted by WL1
and WL2. WL1 and WL2 are defined as the regions which are respectively
within a distance of $6.25$~\AA~and $3.0$~\AA~away from the wall
corresponding to the second minimum of the density profile for silicon
and the first minimum of the total density profile, respectively (see
Fig.~\ref{fig2}a). Note that WL2 is identical with WL. In each region,
i.e.~bulk, WL1, and WL2, we took only those rings into account that fit
completely into it. Thus, in WL2 those rings are counted that are formed
at each case by the first and the second O and Si layers (with respect
to the distance from the wall), whereas with WL2 only those rings are
taken into account that are formed by the first O and the first Si layer.
This is justified because the first two oxygen and the first two silicon
layers are well--defined in that the minima in the corresponding density
profiles are close to zero density in the case of WL2 and around the
small value $\rho=0.5$~g$/$cm$^{3}$ in the case of WL1. Furthermore,
we can infer from Fig.~\ref{fig2}a that in contrast to the second layers
the first oxygen layer overlaps strongly with the first silicon layer and
the overall thickness of both layers is only about $2$~\AA.  Thus, the
first oxygen and the first silicon layer form a quasi--two--dimensional
plane and $P(n)$ for WL2 gives a distribution of rings that have an
orientation parallel to the walls whereas in $P(n)$ for WL1 also the
rings perpendicular to the walls are included.

As has been also found in pure bulk simulations of
SiO$_2$~\cite{vollmayr96}, in the bulk a maximum is observed around $n=6$.
This is plausible since in silica the high--temperature crystalline phase
at zero pressure, $\beta$--cristobalite, exhibits only six--membered
rings.  In WL1 the probability for $n\ge 6$ is smaller than in the bulk
in favor of a relatively high probability of $n=3$ and $n=4$. In WL2 it
is the other way round: $n=4$ and even more $n=5$ are less frequent than
in the bulk in favor of $n=8,9,10$. The ring structure near the walls
that corresponds to these findings is as follows: Perpendicular to the
walls small rings with $n=3,4$ are seen such that, e.g., $n=3$ is formed
by two silicon atoms from the first silicon layer with a third one from
the second silicon layer (for an illustration see Fig.~\ref{fig1}). In
contrast to that, parallel to the walls (considering the first oxygen
and the first silicon layer) an open structure with relatively large
rings is observed which compensates somewhat the dense packing of SiO$_4$
tetrahedra perpendicular to the walls.

Fig.~\ref{fig5}b shows the behavior of $P(n)$ in the bulk at the
two temperatures $T=3760$~K and $T=5200$~K in equilibrium and under
shear. We can immediately infer from the figure that the considered
shear fields have only a small effect on the structure which confirms
the findings for all the quantities discussed before. At $T=5200$~K one
has a relatively large amount of two-- and three--membered rings the
frequency of which is more than a factor of two smaller at $T=3760$~K. In
Ref.~\cite{vollmayr96} it was shown that their frequency of occurrence
decreases further with decreasing temperature such that the amount
of two--membered rings falls far below 1\% for systems that have typical
structural relaxation times of the order of $1$~ns.

In contrast to the bulk in WL1 significant changes in the ring
distribution take place if the system is sheared (see Fig.~\ref{fig5}c),
and the external force field affects the ring structure such that small
and large rings are formed, while at the same time especially the amount
of six--membered rings decreases. Only for the smaller field strength
$a_{{\rm e}}=3.8$~\AA$/$ps$^2$ there are no significant changes in the
ring distribution. We will see below that the latter is accompanied with
a very small slip motion at the walls whereas a large slip velocity is
correlated with strong rearrangements in the ring structure.  One can
also infer the remarkable fact from Fig.~\ref{fig5}c that the probability
to find rings with $n=3,4$ does not change very much when an external
acceleration field is switched on. This is reasonable because these
small--membered rings, as we have seen before, are located perpendicular
to the walls and thus they are very stable to shear forces that are
imposed parallel to the walls.

The strongest rearrangements in the ring structure due to a shear field
are found when we consider the region WL2. The corresponding curves are
shown in Fig.~\ref{fig5}d. Again, there are only minor changes in $P(n)$
at $T=5200$~K and $a_{{\rm e}}=3.8$~\AA$/$ps$^2$ as compared to the
corresponding equilibrium case which is, as mentioned before, related
to the presence of only a very small slip velocity. For the higher
acceleration field, $a_{{\rm e}}=9.6$~\AA$/$ps$^2$, the ring structure
becomes more heterogeneous and the effect of the external field is if one
would locally increase the temperature. The rearrangements in the ring
structure can be summarized as follows: Rings mainly of size $n=6,7,8$ are
broken under the influence of the shear force and instead small rings with
$n<4$ and very large rings with $n\ge 9$ are formed.  This is illustrated
in Fig.~\ref{fig6} where snapshots at $T=3760$~K of the WL2 region are
shown for the equilibrium case and the non--equilibrium steady state case
with the external field of strength $a_{{\rm e}}=9.6$~\AA$/$ps$^2$. One
can clearly identify, e.g., rings with $n=2$ and $n=11$ in the right
panel which are absent in the left panel, i.e.~at equilibrium.

Up to now we have seen that in the bulk region the structure essentially
does not change for acceleration fields $a_{{\rm e}} \le 9.6$~\AA$/$ps$^2$
whereas significant structural rearrangements occur close to the walls
for $a_{{\rm e}} = 9.6$~\AA$/$ps$^2$.  We demonstrate now how this is
reflected in the flow properties, and we discuss to what extent the flow
can be understood within a hydrodynamic description.

Fig.~\ref{fig7} shows velocity profiles $v_x(z)$ across the film
for two different choices of the acceleration in $x$--direction for
$T=5200$~K. The solid lines are fits according to the formula
\begin{equation}
   v_x(z) = \frac{\rho a_{{\rm e}}}{2 \eta} (z-z_1)(L_z-z_1-z)
   \label{velprof}
\end{equation}
with $\rho$ being the total mass density, and the intercept $z_1$
as well as the shear viscosity $\eta$ are treated as adjustable
parameters. Eq.~(\ref{velprof}) is the formula for a one--component fluid,
but it is justified to use this equation for our two--component system
since the velocity profiles for the individual species, i.e.~for silicon
and oxygen, are identical within the statistical errors. Note that this
was also observed by Koplik and Banavar~\cite{koplik} for a mixture of two simple
liquids confined between two walls. The values of $\eta$ that result
from the fits with Eq.~(\ref{velprof}) are quoted in the figure. As it
should be, $\eta$ is independent of the acceleration $a_{{\rm e}}$ within
statistical errors. This matches well to the aforementioned result that
the structure is essentially not affected in the bulk by the considered
shear fields. Furthermore, this analysis shows that hydrodynamics
holds down to nanoscopic scales, even if we use a realistic model of
a molecular glassforming fluid such as SiO$_2$. Close to the wall the
fit with Eq.~(\ref{velprof}) is obviously not valid for $a_{{\rm e}}
= 9.6$~\AA$/$ps$^2$ and we find a slip motion of the fluid. As we have
seen before this slip is accompanied with structural rearrangements in
the ring distribution. 

It is conspicuous in Fig.~\ref{fig7} that there is a sharp peak in
$v_x(z)$ at $z=1.2$~\AA~in the curve for $a_{{\rm e}}=3.8$~\AA$/$ps$^2$
(there is exactly the same peak at $z=30.3$~\AA~because the profile is
symmetrized).  This is just due to the fact that there is a very small
probability for the particles to move into the region $z<1.5$~\AA,
i.e.~the region very close to the wall.  And if a particle enters
this region there is a high probability that it has a relatively
high velocity. Of course, on average this velocity should be zero,
but in this case the statistics is bad since this event is very rare,
and we observe sharp peaks in the regions very close to the walls that
can have a negative or positive amplitude with equal probability (note
that they appear also in the unsheared system). A similar reasoning
might be also true at $z=3$~\AA, i.e.~at the location of the 
first minimum in the total density profile.
Indeed we see a small peak at $z=3$~\AA~in the curve for $a_{{\rm e}}=9.6$~\AA$/$ps$^2$
which we also find at lower temperatures
(see Fig.~\ref{fig8}). But we cannot conclude from our data whether this
feature would disappear with a better statistics.

One still could argue that the fits with Eq.~(\ref{velprof}) are
fortuitous, and therefore the resulting shear viscosity $\eta$ then
simply is a fit parameter with little physical significance, different
from the actual shear viscosity of bulk molten SiO$_2$. Fig.~\ref{fig8}
shows that this objection is not true: here the velocity profiles are
compared at three temperatures, and the viscosities $\eta$ from the
fits are compared to viscosities $\eta_{{\rm GK}}$ obtained earlier
from simulations of bulk SiO$_2$ at rest (no walls, no flow) applying
Green--Kubo formulas~\cite{horbach99}. Note that the latter approach
typically involves a relative error of about 20\%, and so does the
present approach. Within these errors, there is a good agreement,
although $\eta_{{\rm GK}}$ changes for the shown range of temperatures
by an order of magnitude.  However, the estimated $\eta$ seems to be
systematically smaller than $\eta_{{\rm GK}}$ at low temperatures and
we cannot exclude that these deviations are due to the influence of
the Nos\'e--Hoover thermostat on the flow velocity.  Of course, we can
also infer from Fig.~\ref{fig8} that we cannot use our simulation of
Poisseuille flow to estimate the viscosity at much lower temperatures
than $3760$~K: If $\eta$ decreases the velocity profile becomes flatter,
and one cannot compensate this by higher values of $a_{{\rm e}}$ since
the slip velocity increases with $a_{{\rm e}}$ also.

\section{Summary and discussion}
\label{sec4}
We used molecular dynamics computer simulations to study a silica melt
confined between two walls. The interactions between the silicon and
oxygen atoms were modelled by the so--called BKS potential for which
it has been shown that it describes reliably the bulk as well as free
surface properties of silica. The walls consist of particles fixed on
a fcc lattice, and wall--fluid interactions were modelled by a simple
12--10 potential. The aim of this work was to investigate how the walls
affect the structure of the melt in equilibrium and under shear, and
whether it is possible to understand the flow properties of the sheared
melt within hydrodynamics.

\begin{figure}[h]
\vspace*{-12cm}
\psfig{file=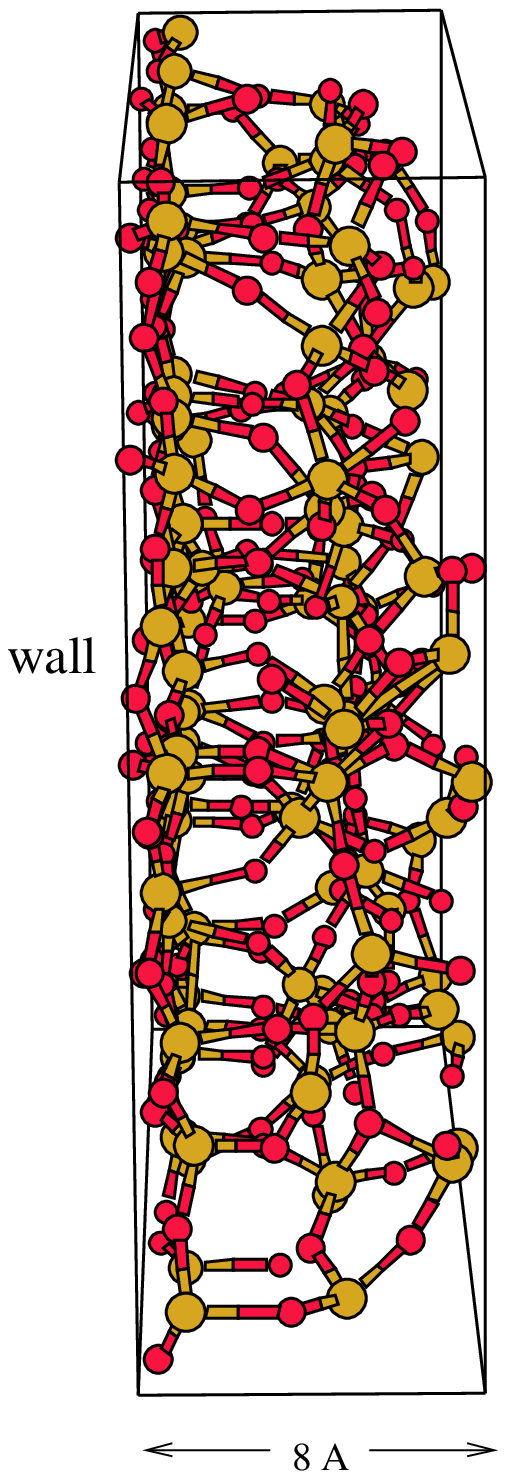,height=25cm}
\caption{Snapshot of a part of a configuration at \protect$T=3760$~K that is within
8~\AA~away from the wall. The brown and the red spheres correspond to the silicon
and oxygen atoms, respectively.
} 
\label{fig1}
\end{figure}

\begin{figure}[h]
\psfig{file=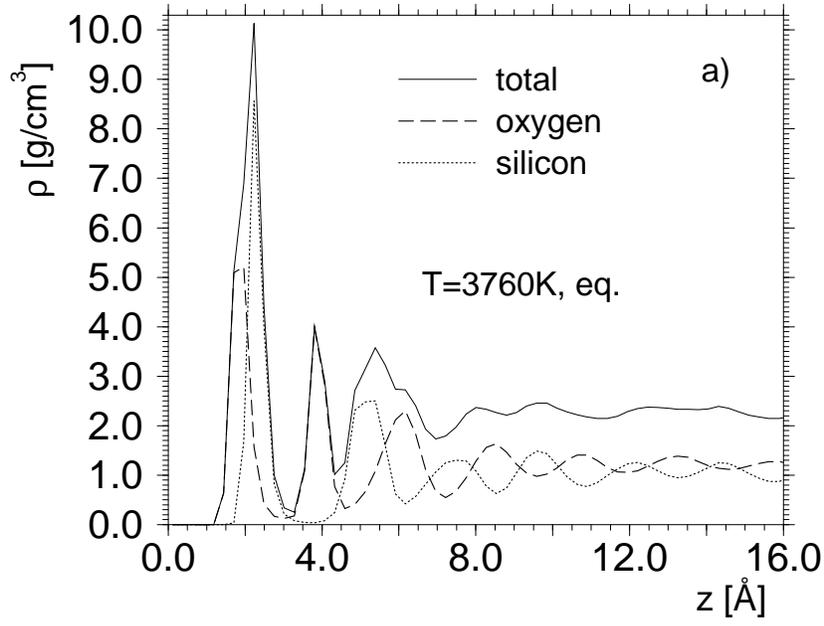,height=8cm}
\vspace*{1cm}
\psfig{file=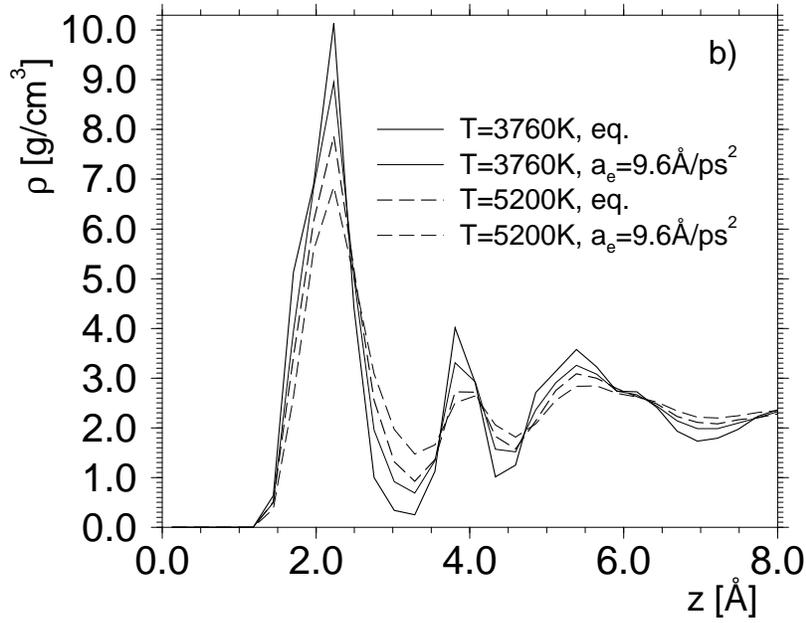,height=8cm}
\vspace*{1cm}
\caption{a) Total density profile and partial density profiles for oxygen and silicon at
            \protect$T=3760$~K.
         b) Total density profiles at equilibrium and with acceleration field 
            \protect$a_{{\rm e}} = 9.6$~\AA$/$ps$^2$ for the two temperatures
            \protect$T=3760$~K and \protect$T=5200$~K.
}
\label{fig2}
\end{figure}

\begin{figure}[h]
\psfig{file=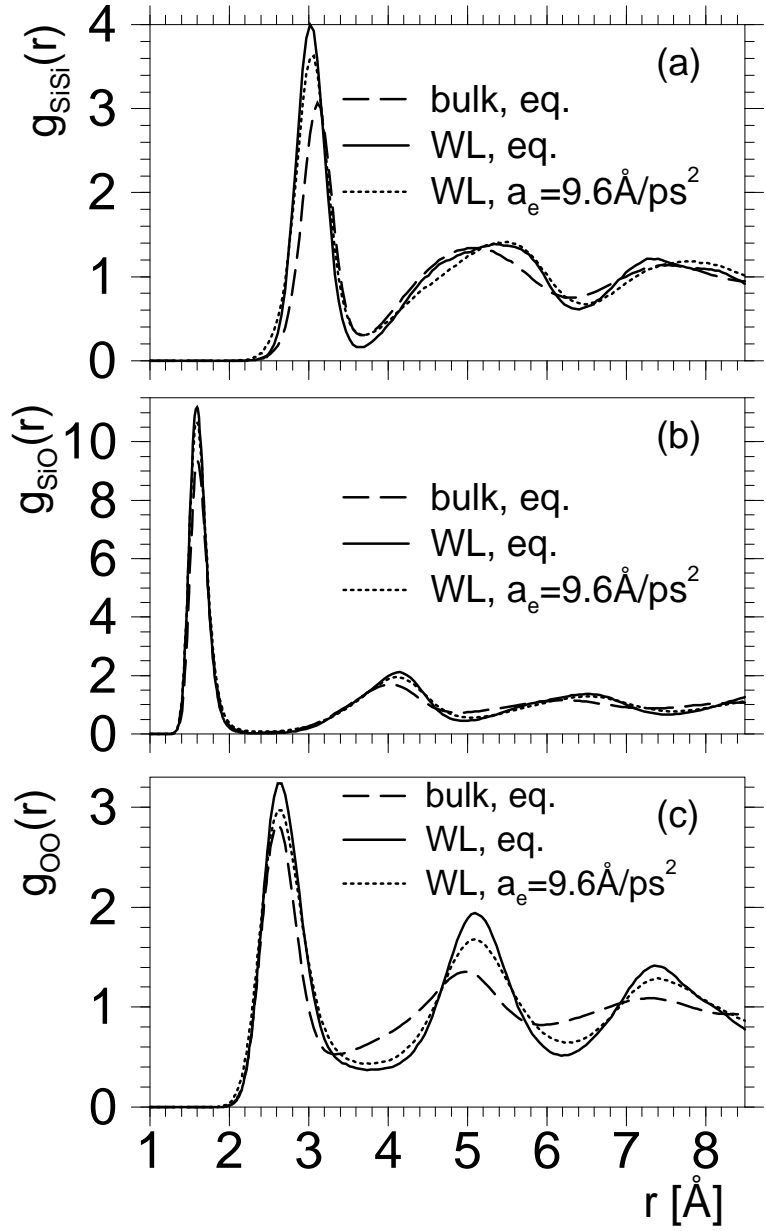,height=16cm}
\vspace*{1cm}
\caption{Partial radial distribution functions \protect$g_{\alpha \beta}(r)$ at equilibrium
         in the bulk and in the WL, and with the field \protect$a_{{\rm e}} = 9.6$~\AA$/$ps$^2$ 
         in the WL. The temperature is \protect$T=3760$~K. a) \protect$g_{{\rm SiSi}}$, 
         b) \protect$g_{{\rm SiO}}$, c) \protect$g_{{\rm OO}}$. 
}
\label{fig3}
\end{figure}

\begin{figure}[h]
\psfig{file=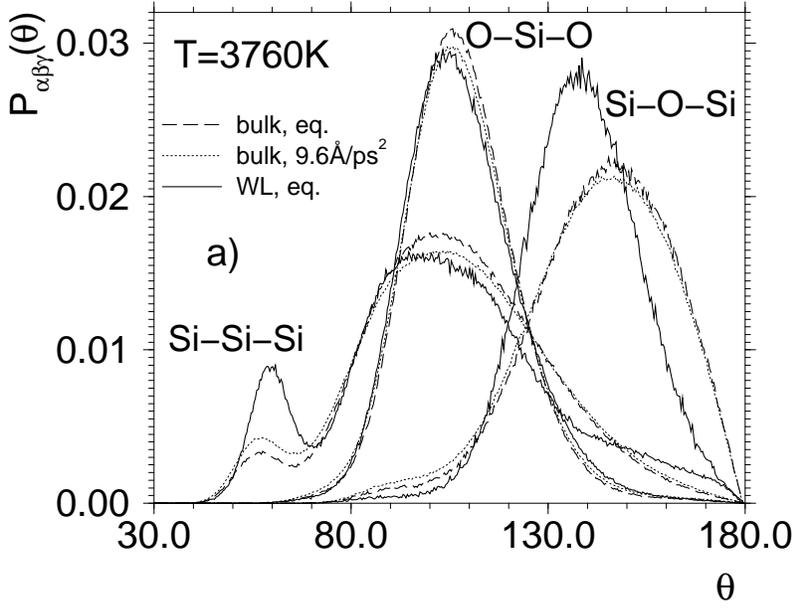,height=8cm}
\vspace*{1cm}
\psfig{file=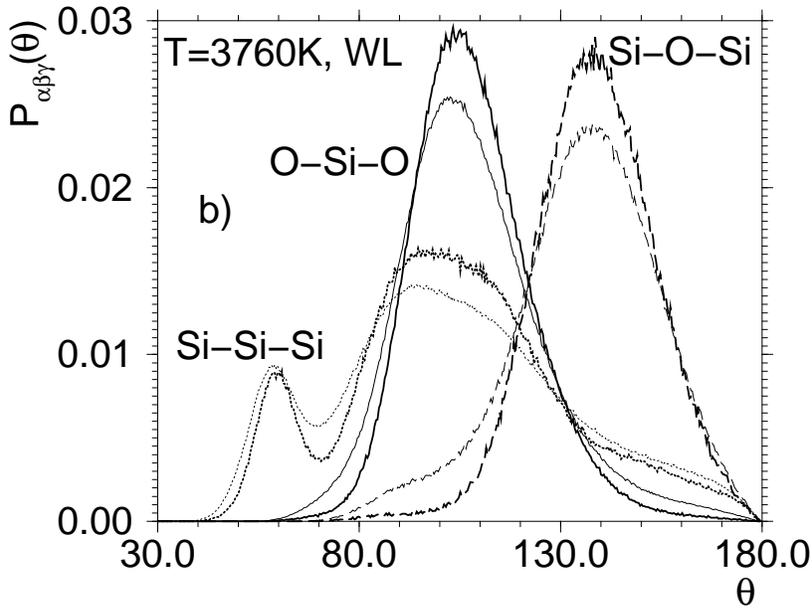,height=8cm}
\vspace*{1cm}
\caption{Distributions of the indicated angles at equilibrium and with external
         field \protect$a_{{\rm e}} = 9.6$~\AA$/$ps$^2$ for \protect$T=3760$~K.
         a) Bulk and WL (at equilibrium), b) WL at equilibrium (bold lines) and
         with external field (thin lines). 
}
\label{fig4}
\end{figure}

\begin{figure}[h]
\psfig{file=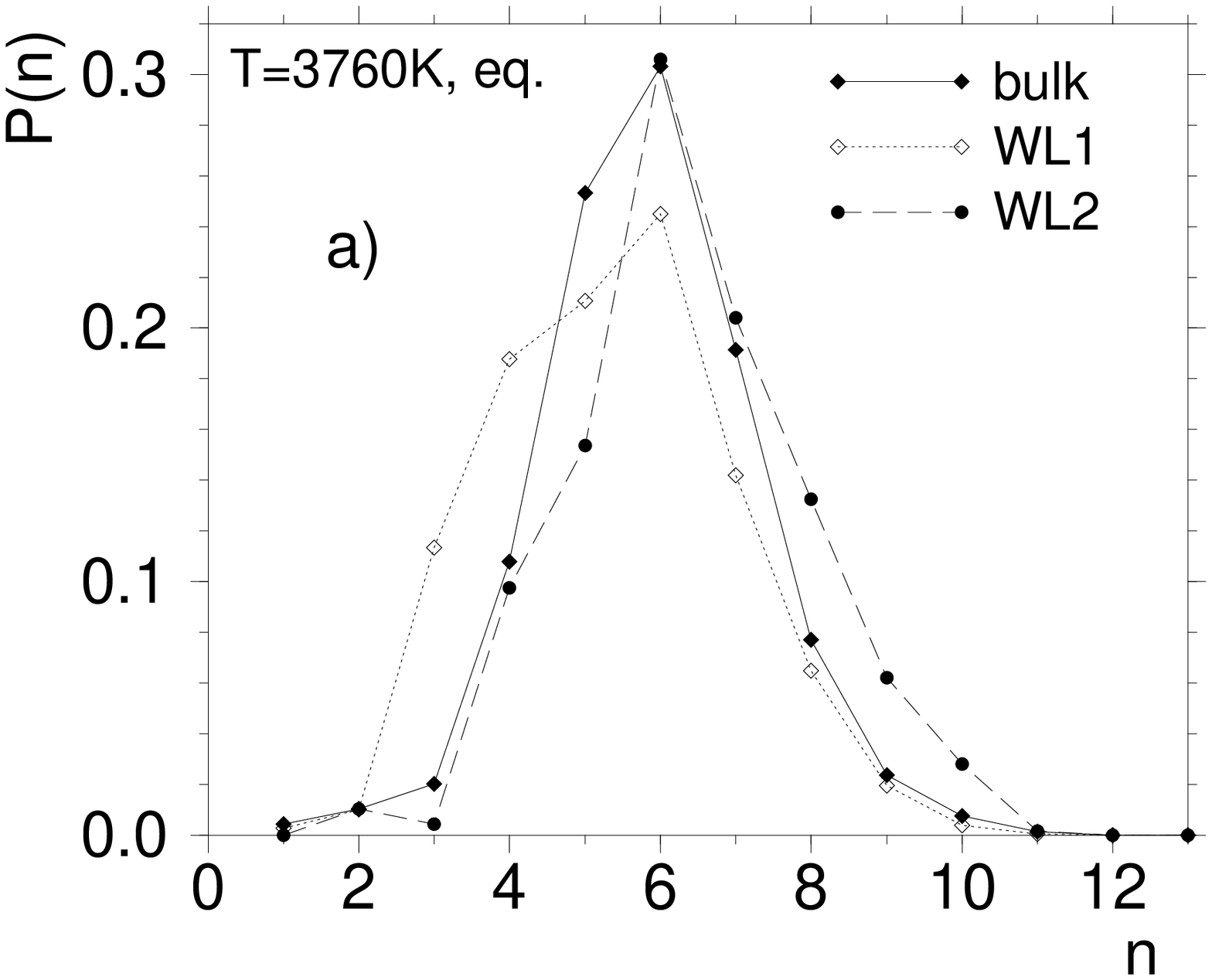,height=8cm}
\vspace*{1cm}
\psfig{file=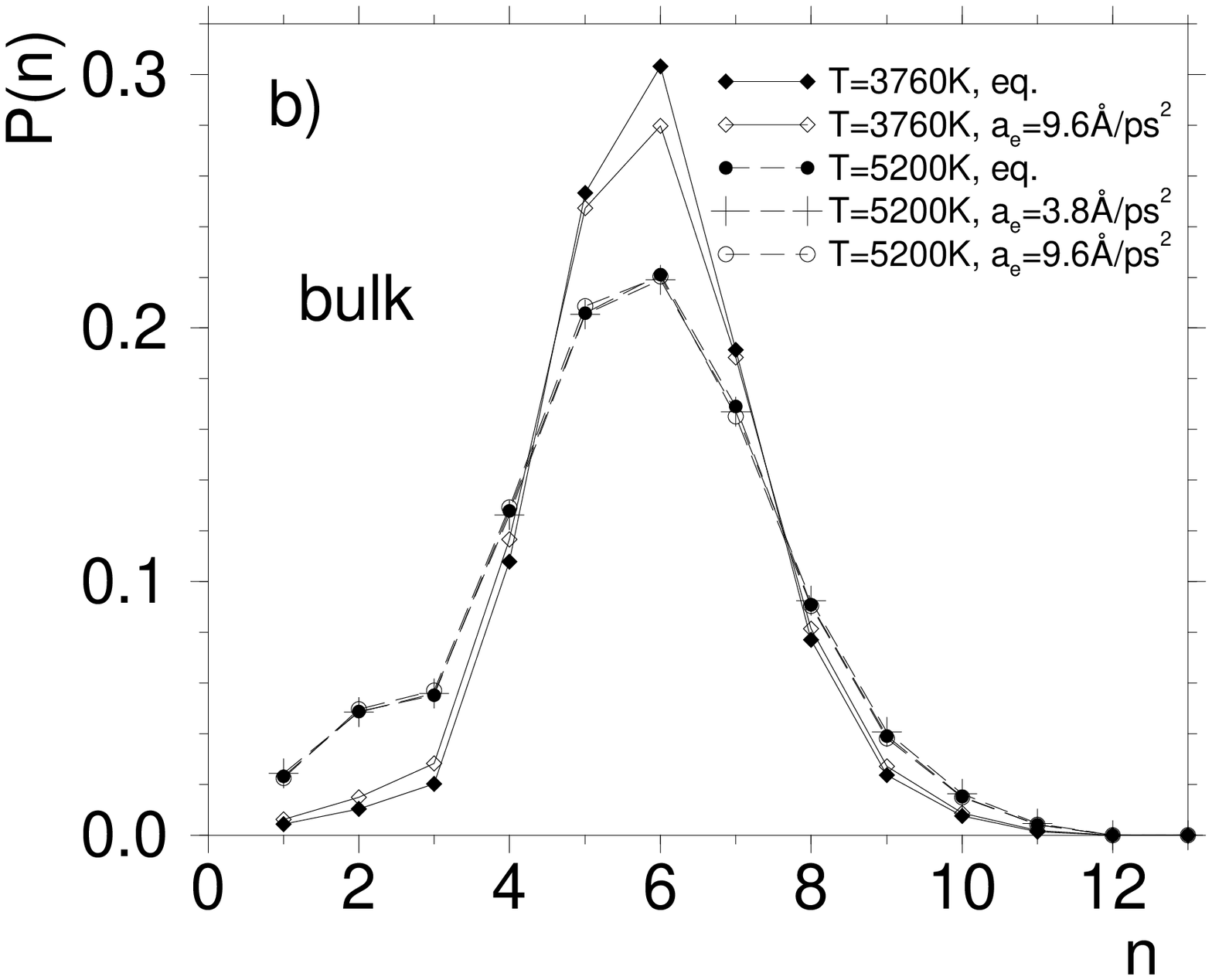,height=8cm}
\psfig{file=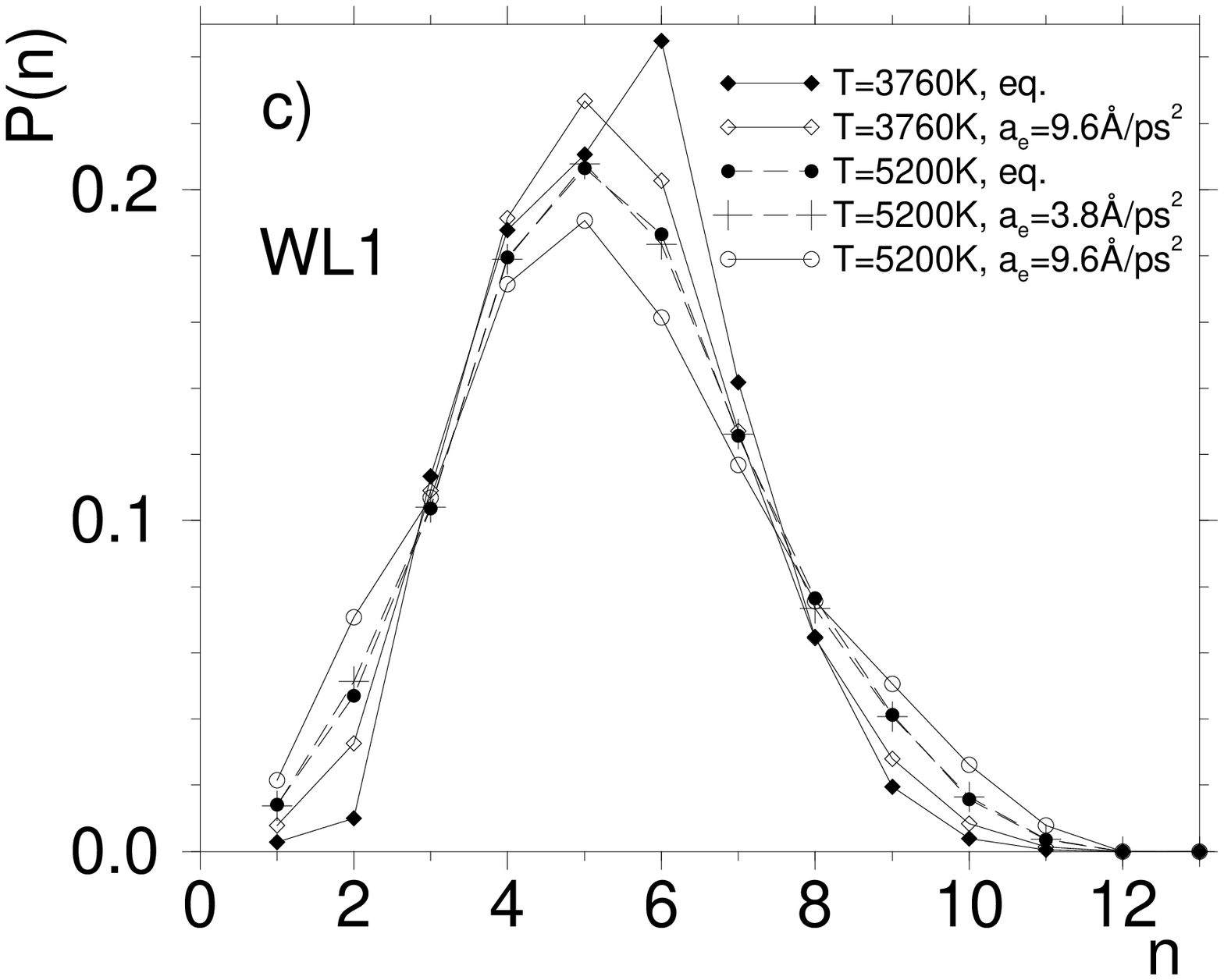,height=8cm}
\vspace*{1cm}
\psfig{file=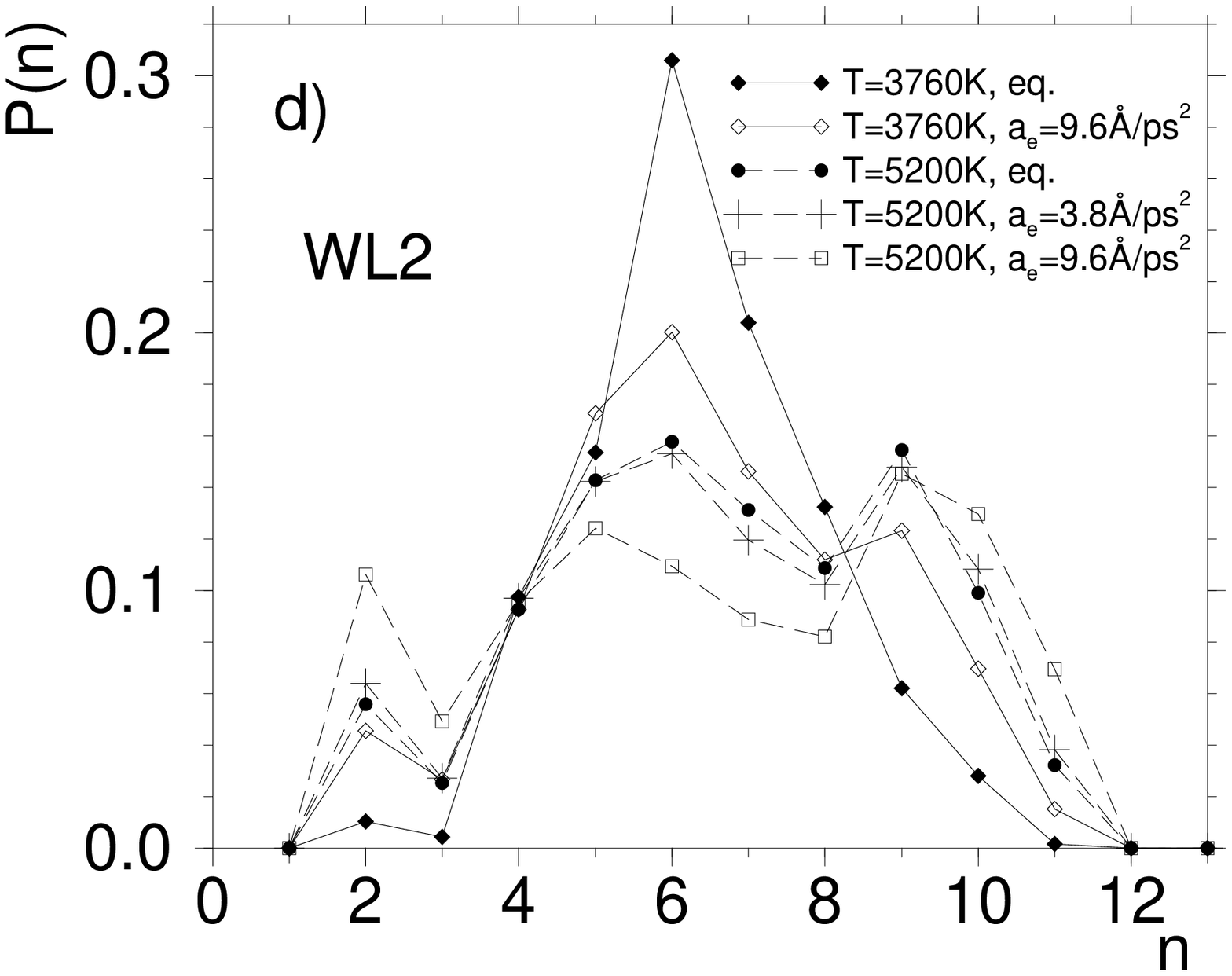,height=8cm}
\vspace*{1cm}
\caption{Probability \protect$P(n)$ that a ring has a length \protect$n$,
         a) at \protect$T=3760$~K in equilibrium for the bulk, WL1, and WL2,
         b) at \protect$T=3760$~K and \protect$T=5200$~K in equilibrium
         and under the indicated acceleration fields in the bulk,
         c) the same as in b) but for WL1, and
         d) also the same as in b) but for WL2.
         For the definition of WL1 and WL2 see text.
}
\label{fig5}
\end{figure}

\begin{figure}[h]
\vspace*{-8cm}
\hspace*{-9cm} \psfig{file=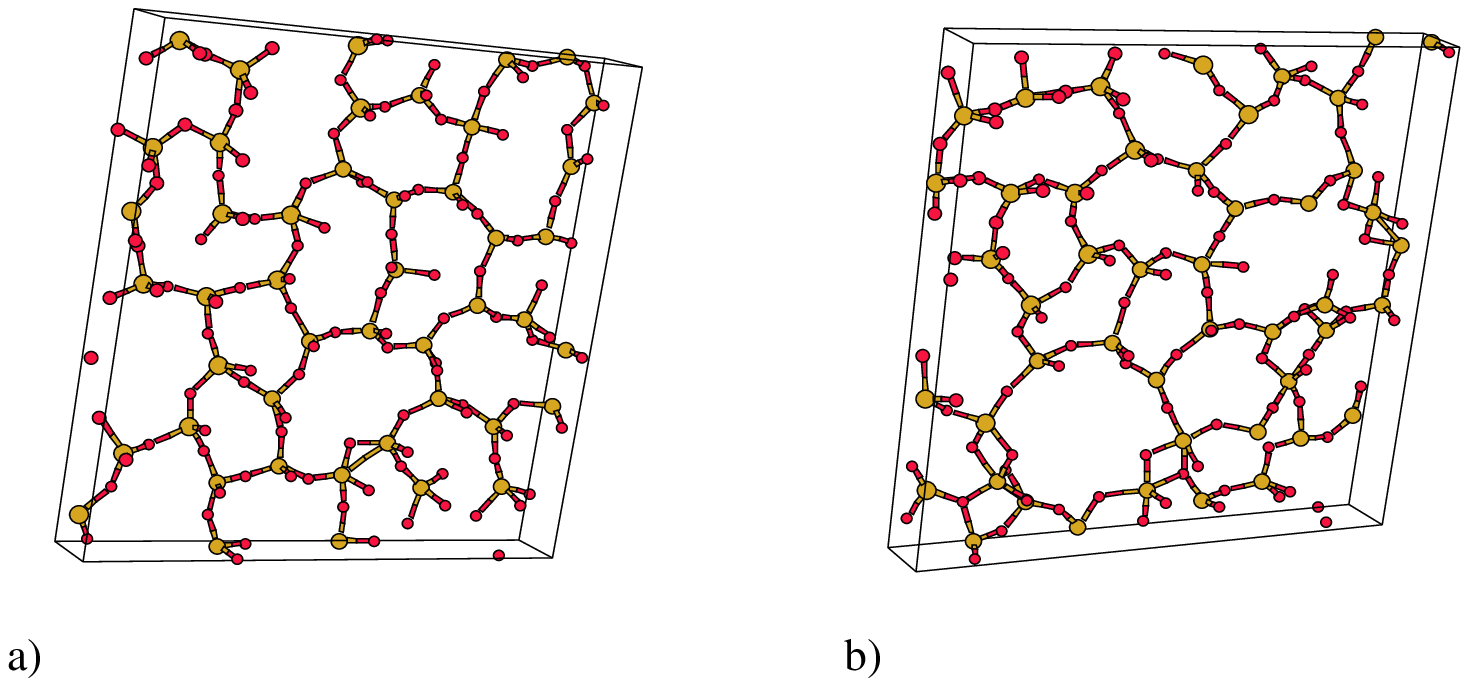,height=25cm,angle=270}
\vspace*{-8cm}
\caption{Snapshot of the wall layers (WL2) for \protect$T=3760$~K, a) at equilibrium and
         b) with the acceleration field \protect$a_{{\rm e}} = 9.6$~\AA$/$ps$^2$ (the direction of
         $a_{{\rm e}}$ is from the left to the right). The dark
         and the light grey spheres correspond to the silicon and oxygen atoms, respectively.
}
\label{fig6}
\end{figure}

\begin{figure}[h]
\psfig{file=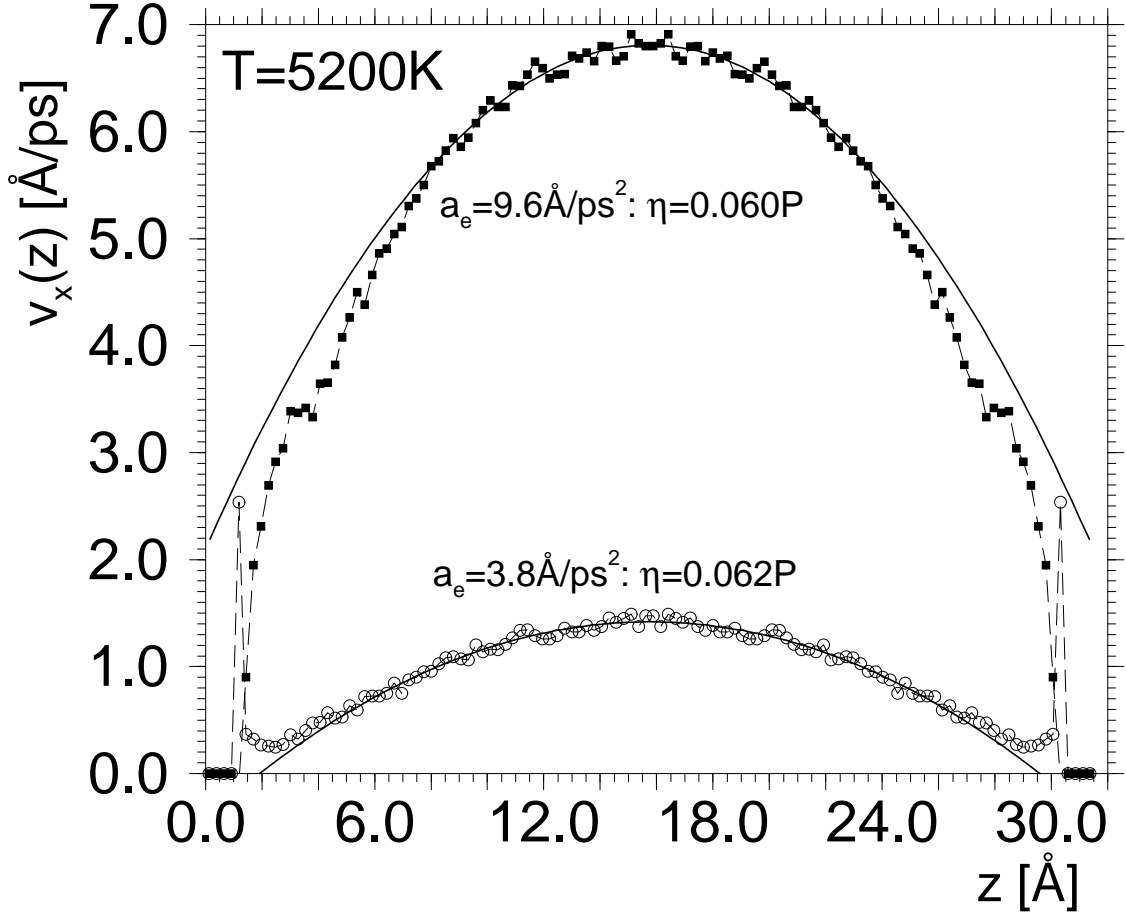,height=12cm}
\vspace*{1cm}
\caption{Velocity profiles at \protect$T=5200$~K for the indicated acceleration fields.
         The bold lines are fits with formula (\protect\ref{velprof}) and the indicated
         values for the shear viscosity \protect$\eta$ are extracted from these fits.
}
\label{fig7}
\end{figure}

\begin{figure}[h]
\psfig{file=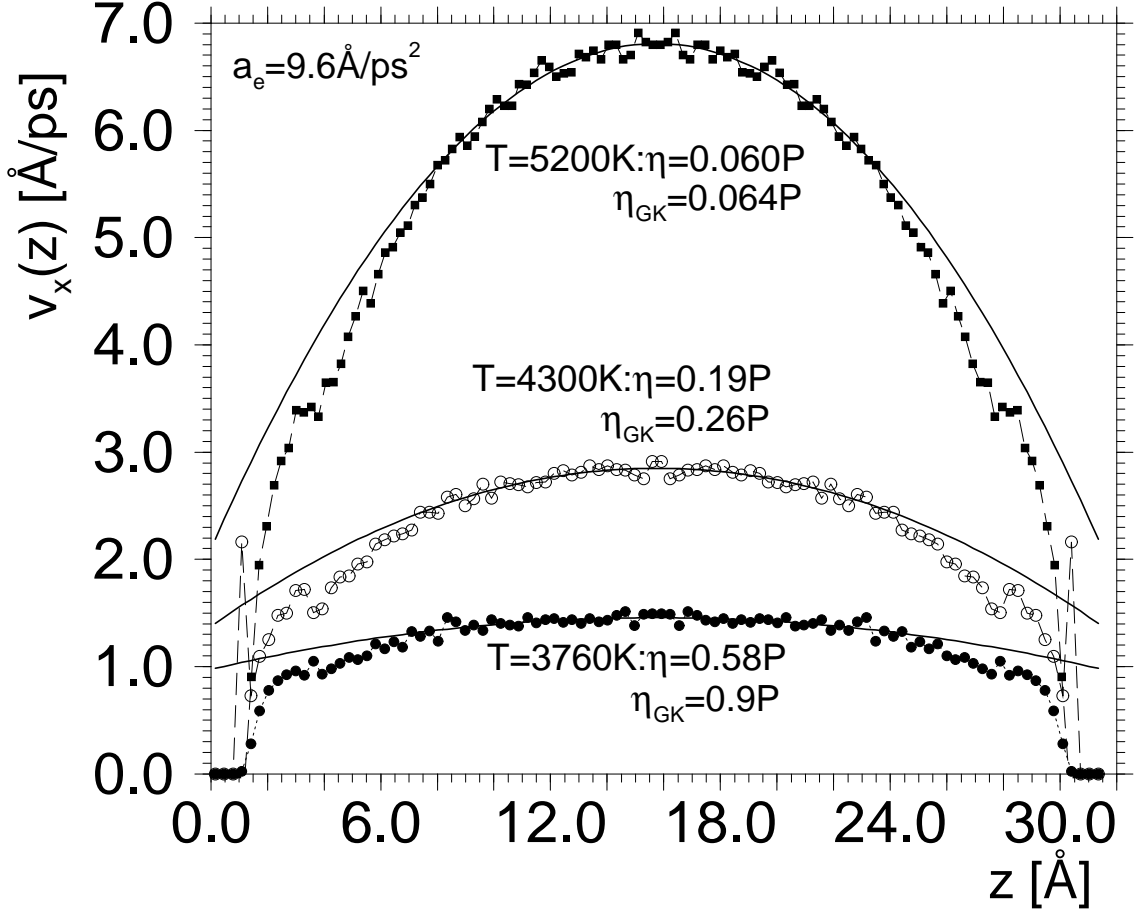,height=12cm}
\vspace*{1cm}
\caption{Velocity profiles for \protect$a_{{\rm e}} = 9.6$~\AA$/$ps$^2$ at different 
         temperatures. \protect$\eta_{{\rm GK}}$ denotes the ``Green--Kubo'' results that are
         taken from Horbach and Kob~\protect\cite{horbach99}. The rest is similar to Fig.~\protect\ref{fig7}.
}
\label{fig8}
\end{figure}

As in the case of simple liquids a pronounced layering of the melt close
to the walls is observed. This layering, as seen in the total density
profile, has a very irregular character and this is related to the specific
chemical ordering of silicon and oxygen atoms in a tetrahedral network.
In the layers that are closest to the walls the tetrahedra prefer an
orientation such that three oxygen atoms of each tetrahedron form the
first layer, the silicon atoms the second one and the remaining fourth
oxygen atoms of the tetrahedra the third one (see Fig.~\ref{fig1}).

It has been demonstrated in recent simulation studies of free silica
surfaces~\cite{roder01,mischler02} that the ring distribution function
is a quantity that is well suited to characterize the differences
of the surface structure on intermediate length scales from that in
the bulk. This holds also true in the present case: The relatively
high probability of small and large rings makes the difference of the
melt structure at the wall from that of the bulk. Perpendicular to the
first wall layer there is a high probability of rings of size $n=3,4$
whereby, e.g., a ring with $n=3$ is typically formed by two silicon atoms
from the first silicon layer and one from the second one (of course,
these rings are also decorated with oxygen atoms). Parallel to the walls,
at least for temperatures $T\le 3760$~K a relatively high probability
for rings of length $n\ge 9$ is found in the wall layers which can be
seen as the complement to the occurrence of small--membered rings with
$n=3,4$. Our findings for the structure of the silica melt near the wall
can be expected to be generic if the wall--SiO$_2$ interactions are,
as in our case, relatively weak, i.e.~no covalent bonds of the wall
atoms with those of the fluid are formed as it would be the case, e.g.,
for an interface between crystalline and amorphous silica.

Under shear, the bulk structure of the silica melt does not change for the
field strengths considered in this work, $a_{{\rm e}} \le 9.6$~\AA$/$ps$^2$. This
is an indication that it is justified to treat the melt as a Newtonian fluid.
At the walls, the behavior of the fluid is different: If a slip motion occurs
there are strong rearrangements in the ring distribution as compared to the 
equilibrium case. Rings of medium size break and, as a result of this, small
and large rings are formed (the same effect is achieved by a local increase of the
temperature). Thus, we see a rather unusual mechanism for slip motion: Slip
seems to occur only if the external (gravitational--like) force field is strong
enough to break rings of medium length. This slip mechanism can be expected to
be generic for liquids with a tetrahedral order (of course, as before, only
if walls are present that do not have a (strong) covalent bonding with the 
particles in the tetrahedral network).

As expected from hydrodynamic theory of a Newtonian fluid we find
parabolic velocity profiles in the bulk region, although the latter region
extends in the systems considered in this work only over about 15~\AA. The
values of the shear viscosity $\eta$ that we have extracted from the
profiles are in fair agreement with the ``Green--Kubo'' values from a
bulk simulation. So the question arises whether the non--equilibrium
molecular dynamics technique used in this work offers a method
to estimate transport coefficients. It has been demonstrated for
a lattice gas model~\cite{kehr89} that non--equilibrium simulation
techniques can be very powerful to gain insight into the understanding of
transport processes in binary mixtures. We have shown in this work
that with the present simulations this is possible for melts with a
viscosity up to about 1~Poise. For systems with higher viscosities
(at lower temperatures) a different wall model would be helpful with
which the tendency for slip motion is smaller and furthermore, larger
system sizes would be required. But for larger system sizes one needs
for the computation of the quasi--two--dimensional Coulomb forces faster
methods than the Parry--Ewald method used in this work. A promising
new method~\cite{arnold02,joannis02} (also based on Ewald sums) that
scales with the number of particles like $N \log N$ has been proposed
very recently and could be applied to systems like the one considered in this
work.

Acknowledgments:
The authors are grateful to W. Kob for very valuable discussions and ideas.
We also thank F. Varnik and M. M\"user for constructive hints and C. Mischler
for a critical reading of the manuscript. Furthermore we
acknowledge partial financial support from the D.I.P. Project 352--101 and
we thank the HLRZ Stuttgart for a generous grant of computer time on the 
Cray T3E.

\begin{references}
\bibitem{foot1}
Here and in the following only symmetrized profiles are shown.
\end{references}
\end{document}